\documentstyle[prd,aps,epsf,epsfig]{revtex}
\begin{document}
\draft
\twocolumn[\hsize\textwidth\columnwidth\hsize\csname@twocolumnfalse\endcsname

\title{Anisotropic finite-size scaling analysis of a
three-dimensional driven-diffusive system}

\author{Kwan-tai Leung$^\dagger$ and Jian-Sheng Wang$^*$}

\address{
$^\dagger$Institute of Physics, Academia Sinica,
Nankang, Taipei 11529, Taiwan, Republic of China\\
$^*$Department of Computational Science, 
National University of Singapore, Singapore 119260,
Republic of Singapore
}

\maketitle
\centerline{\small (20 May 1998,  revised 29 December 1998)}

\begin{abstract}

We study the standard three-dimensional driven diffusive system
on a simple cubic lattice where particle jumps along a given
lattice direction are biased by an infinitely strong field, while
those along other directions follow the usual Kawasaki dynamics.
Our goal is to determine which of the several existing theories
for critical behavior is valid.
We analyze finite-size scaling properties
using a range of system
shapes and sizes far exceeding previous studies.  
Four different analytic predictions are tested against the numerical
data. Binder and Wang's prediction does not fit the data well.
Among the two slightly different versions of Leung, the one
including the effects of a dangerous irrelevant variable appears
to be better.  Recently proposed
isotropic finite-size scaling is inconsistent with our
data from cubic systems, where systematic deviations are found, 
especially in scaling at the critical temperature. 

\end{abstract}

\pacs{PACS number(s): 05.50.+q, 64.60.C, 05.70.Jk.}
\vspace{2pc}
]

\section{Introduction}

Over the last decade or so, there have been many investigations
on nonequilibrium systems.  One of the most often studied is the
driven diffusive system (DDS) \cite{review} for its simplicity
of formulation and richness of novel properties. In particular,
it is one of the few simple nonequilibrium systems showing phase
transitions similar to that of an equilibrium
statistical-mechanical system.  The distinction between DDS and
an equilibrium system is a subtle one.  Although both reach a
stationary state of their respective stochastic dynamics in the
long time limit, the DDS is generically nonequilibrium, defined
only through its dynamics, while an equilibrium system can be
alternatively characterized by a Hamiltonian independent of the
choice of dynamics.  The {\em standard\/} DDS model is a lattice
gas model governed by Kawasaki dynamics with a driving field.
The driving field biases the motion of the particles in one
preferred direction, so that under periodic boundary conditions
it gives rise to a current along that direction.  The existence
of a steady current is a manifestation of the nonequilibrium
nature of the stationary state.  One physical realization of DDS
is the superionic conductor.  Certain flow properties of binary
liquids under gravity, though more complicated due to
hydrodynamic modes, are also similar to those of DDS.
 
In this paper we focus our attention on the critical behavior of
the standard DDS.  Mean-field theories \cite{mean-field} can
give qualitative predictions of the phase transitions.  But one
of the achievements of the field-theoretic treatment of DDS
\cite{Janssen86,LeungCardy86} is the exact determination of 
the set of critical exponents for all dimensions $d
\ge 2$.  After some controversies, the predictions in two dimensions
have been confirmed by extensive Monte Carlo simulations
\cite{Leungprl91,Wang96}.  The most difficult aspect of such 
tests is the fact that two length scales are involved---the
correlation lengths ($\xi_\parallel$ and $\xi_\perp$) diverge
differently in direction parallel to the field and in direction
perpendicular to the field.  Thus, one has to deal with
anisotropic finite-size scaling with systems of various
geometries.

From a field-theoretic point of view, the $d=2$ model is more
complicated than in higher dimensions. This is because the usual
$\phi^4$ coupling constant, denoted by $u$, is a dangerous
irrelevant variable for $d>2$, but it becomes marginal in $d=2$.
While scaling arguments can predict the effect of $u$ on
finite-size scaling for $d>2$ \cite{Leungprl91}, there is little
clue as to the existence and possible form of the associated
logarithmic corrections in $d=2$.  Perhaps this explains why the
agreements between previous tests\cite{Leungprl91,Wang96} and
theory in $d=2$ are not impeccable; small deviations from
scaling may be due to the presence of small logarithmic
corrections.  For this reason, a more stringent yet practical
test lies in $d=3$.

Recently, not only the confirmations in $d=2$
\cite{Leungprl91,Wang96} but also the validity of the
field-theoretic approach itself have been questioned by Marro et
al \cite{Marro96}.  
Isotropic finite-size scaling involving one scaling length was advocated.
It is therefore our aim here to try to
answer those questions by conducting a comprehensive test of the
field-theoretic predictions in $d=3$.  Being free from the doubt
of possible logarithmic correction, consistencies between
simulations and field theory would lend strong support to the
latter.

There are very old Monte Carlo simulation results
\cite{Marro87,Zhang88} of the three-dimensional DDS.  One of 
the discoveries by Monte Carlo as well as by analytic works is
the power-law long-range correlation of the particle density
even in the high-temperature disordered phase
\cite{review,Zhang88,Garrido90}.  This is shown as the manifestation
of the violation of the fluctuation-dissipation theorem in DDS.
But the question of the validity of the field-theoretic results
is far from answered.  Monte Carlo simulation of the DDS model
is difficult for its long relaxation times and anisotropic
correlations, due to local conservation and the external drive,
respectively.  In this article, we report a fairly extensive
Monte Carlo study using anisotropic finite-size scaling analysis
similar to that in the two-dimensional case.  A dominant feature
of the anisotropies is the appearance of an extra scaling
variable, the ``aspect ratio'' $S =
L_{\parallel}^{1/\lambda}/L_{\perp}$, with
$\lambda=\nu_\parallel/\nu_\perp$.  The exponents $\nu_\perp$ and
$\nu_\parallel$ are associated with the
correlation lengths $\xi_\perp$ and $\xi_\parallel$.  Simple
data collapse among different samples is possible only when
they have the same value of $S$, then the ordinary finite-size
scaling forms are valid.  There have been different predictions
for the value of $\lambda$: In $d=3$, field theory predicts that
$\lambda=8/3$, whereas Binder and Wang \cite{Binder89} obtained
$\lambda=4$.  In this work, we mainly concentrate on samples
with one fixed $S$, using the value 8/3 for $\lambda$.  The
results support the field-theoretic predictions.  Due to the
huge demand on computer power, we could not study extensively
the dependence on $S$, and the scaling at $T_c$, but only have
some limited checks.
We also test the assumption of isotropic scaling
which corresponds to $\lambda=1$.  Inconsistencies are found.

\section{The DDS Model and Simulation Technique}
\label{sec:model}

The model is defined on a simple cubic, fully periodic lattice
of size $L_\parallel \times L_\perp^2$.  Each site on the
lattice has a spin $\sigma_i=\pm 1$.  Equivalently, we can also
consider the system as a lattice gas with local occupation
variables.  The total magnetization is set exactly at zero, and
we assume ferromagnetic interaction among nearest neighbors
only, with a coupling constant $J>0$.  Equivalently, the
particle occupation is half filled and they attract each others.
The system evolves according to the standard Kawasaki dynamics
of spin exchanges except with an extra ingredient due to an
external ``electric'' field.  We associate a positively charged
particle with an up spin and a hole with a down spin.  Particle
hoppings along the electric field are favored.  Hereafter we
will use the spin language.  In simulation, we consider only the
extreme case of infinitely strong field, chosen to be in the
$+x$ direction.  Thus, exchange is always performed if we have a
$(+, -)$ pair along the $+x$ direction and the exchange for
$(-,+)$ is forbidden.  When the exchanges are perpendicular to
the field (which happens 2/3 of the time), the field does not
play any role.  In that case, we compute the change in energy
$\Delta E$ due to the exchange, accepting it with the Metropolis
rate $\min\bigl(1, \exp(-\Delta E/kT)\bigr)$.  From now on, we
will set $J/k$ to one.

Since the model evolves very slowly via a conservative dynamics,
a fast algorithm is of the utmost importance for the present
undertaking.  We used a multi-spin coded program to simulate 8
or 16 systems simultaneously, depending on the word length of
the given machine.  The method is similar to that of Kawashima
et al \cite{multispin}, capable of achieving a speed of about
0.1 $\mu$sec/(spin flip) with a typical workstation.

We define the order parameter as 
\begin{equation}
\phi = {1\over 2 L_\parallel L_\perp} \sin\left({\pi\over L_\perp}\right)\;
\sqrt{ |\tilde \sigma(0,1,0)|^2 + |\tilde \sigma(0,0,1)|^2},
\end{equation}
where
\begin{equation}
\tilde \sigma(l, m, n) = 
 \sum_{x=0}^{L_\parallel-1}\!
\sum_{y=0}^{L_\perp-1}\!\sum_{z=0}^{L_\perp-1}
\!\!\sigma_{x,y,z} e^{2\pi i l x \over L_\parallel } 
e^{2\pi i (m y+ n z) \over L_\perp }.
\end{equation}
The normalization is chosen such that $\phi = 1$ for a slab
geometry (the completely phase-separated configuration in the
limit $T \to 0$).  The following quantities are calculated: 
(i) the averaged order parameter $m = \langle \phi \rangle$,
(ii) the ``susceptibility'' proportional to the fluctuation of
the order parameter,
\begin{equation}
\chi = {L_\parallel L_\perp \over T \sin(\pi/L_\perp)}\; 
   \Bigl[ \langle \phi^2 \rangle
 -  {\langle \phi \rangle}^2 \Bigr],
\end{equation}
the susceptibility above the critical temperature,
\begin{equation}
\chi' = 
{L_\parallel L_\perp\over T \sin(\pi/L_\perp)}\; \langle \phi^2 \rangle, 
\end{equation}
and (iii) the fourth-order cumulant,
\begin{equation}
g=2-{\langle \phi^4 \rangle \over {\langle\phi^2 \rangle}^2}.
\end{equation}
Note that $g$ goes from 0.5 to 1 as temperature $T$ goes from
$\infty$ to 0.  We will not report the results on $\chi'$; it
yields no additional information as $\langle \phi^2 \rangle$
appears to scale like $\langle\phi\rangle^2$.
 
The computations are performed on a variety of workstations,
such as Alpha-stations, Pentium clusters, IBM SP2, etc.  Our
main results are obtained from a set of system sizes
$(L_\parallel, L_\perp)$ with $(113,18)$, $(193,22)$,
$(367,28)$, and $(524,32)$, chosen such that
$S=L_\parallel^{3/8}/L_\perp$ is very close to a constant,
ranging between 0.32703 to 0.32709.  A second set with
$(59,35)$, $(102,43)$, $(122,46)$, $(161,51)$, and $(248,60)$
are used mainly to confirm the result for $T_c$. The value of
$S$ for this set varies from 0.13171 to 0.13182.  Much more
elongated geometries are also used to investigate the
$L_\parallel \to \infty$ behavior.  A third set with cubic
geometry, $L = L_\perp = L_\parallel =$ 20, 30, 40, 50, 60,
is used to test isotropic scaling.  The lengths of runs are
$10^6$ to $10^8$ Monte Carlo steps per temperature, the longer
for $T$ closer to $T_c$.  We monitor the results until the
system is well equilibrated before actually taking data.  The
total amount of CPU time spent is of the order of seven years on
one IBM SP2 node.  This gives an idea of the computational
demand in achieving good statistics for this model.
 
\section{Anisotropic finite-size scaling}
\label{sec:scaling}

There have been two competing theories on the anisotropic
finite-size scaling of the DDS.  The first is that of Binder and
Wang \cite{Binder89}.  Their results are based on a
generalization of the one-dimensional Ginzburg-Landau type
effective Hamiltonian for the very elongated geometry.  The
second is due to Leung based on the field-theoretic formulation
\cite{Leungprl91}.

Encouraged by the finite-size scaling of the standard Ising
model above its upper critical dimension and the finite-size
scaling at a Lifshitz point, Binder and Wang \cite{Binder89}
speculated on a scaling form for the driven diffusive model.
The starting point is the assumption of an effective functional
for the local order parameter $\Psi$ in a quasi-one-dimensional
geometry ($L_\perp \ll L_\parallel$),
\begin{equation}
H_{eff}(\Psi) = L_\perp^{d-1} \! \int_0^{L_\parallel}\!\! dz \left[
{1\over 2} \left( {d^\kappa \Psi \over d z^\kappa} \right)^2\! + {1\over 2} t \Psi^2
+ {u_0 \over 4!} \Psi^4 \right],
\end{equation}
where the exponent $\kappa^{-1} = 2 + (5-d)/3$ which
characterizes the singular term is introduced in such a way that
the correlation-length exponent $\nu_\parallel$ as obtained by
field theory is reproduced at the Gaussian level.  This leads to
the scaling forms for the susceptibility and magnetization in
three dimensions \cite{Binder89} at $T=T_c$
\begin{equation}
\chi(T_c) = L_\parallel^{3/4} 
                \tilde{\chi}(L_\parallel^{1/4}/L_\perp),
\label{eq:BW_chi}
\end{equation}
\begin{equation}
m(T_c) = L_\parallel^{-3/8} 
                \tilde{m}(L_\parallel^{1/4}/L_\perp).
\label{eq:BW_m}
\end{equation}

On the other hand, Leung deduced the off-$T_c$ finite-size
scaling forms by generalizing the exact field-theoretic results
for infinite system sizes to finite sizes \cite{Leungprl91}.
The derivation was based on a combination of the renormalization
group argument which treats $1/L_\perp$ and $1/L_\parallel$ as
two independent relevant variables, and scaling arguments which
assume multiplicative singularities in the limit $u \,
L_\parallel^{-2\theta/\lambda}\to 0$.  Here $\theta$ is
proportional to the anomalous dimension of $u$.  The fact that
$\theta=1-(5-d)/3\geq 0$ for $d\geq 2$ prescribes the dangerous
nature of $u$ above two dimensions.  The main results read
\cite{Leungprl91}
\begin{equation}
\chi(T) = L_\parallel^{7/8} 
                \tilde{\chi}(L_\parallel^{3/8}/L_\perp, t L_\parallel^{7/8}),
\label{eq:leung_chi}
\end{equation}
\begin{equation}
m(T) = L_\parallel^{-7/16} 
                \tilde{m}(L_\parallel^{3/8}/L_\perp, t L_\parallel^{7/8}),
\label{eq:leung_m}
\end{equation}
\begin{equation}
g(T) = \tilde{g}(L_\parallel^{3/8}/L_\perp, t L_\parallel^{7/8}),
\label{eq:leung_g}
\end{equation}
where $t = (T-T_c)/T_c$.
If the contribution from the dangerous irrelevant variable $u$ was ignored,
we would have instead 
\begin{equation}
\chi(T) = L_\parallel^{3/4} 
                \tilde{\chi}(L_\parallel^{3/8}/L_\perp, t L_\parallel^{3/4}),
\label{eq:leung_chiwo}
\end{equation}
\begin{equation}
m(T) = L_\parallel^{-1/2} 
                \tilde{m}(L_\parallel^{3/8}/L_\perp, t L_\parallel^{3/4}),
\label{eq:leung_mwo}
\end{equation}
\begin{equation}
g(T) = \tilde{g}(L_\parallel^{3/8}/L_\perp, t L_\parallel^{3/4}).
\label{eq:leung_gwo}
\end{equation}
The above results imply that the thermodynamic limit has to be
taken carefully.  The field-theoretic results are understood to
correspond to the case where $L_\parallel\propto
L_\perp^\lambda\to\infty$.  Besides this limit, the
quasi-one-dimensional limit $L_\parallel \to
\infty$ with $L_\perp$ held finite is also of interest
(we assume that there exists a unique value of $T_c$ for these
different ways of taking the thermodynamic limit.)  
Since $\chi$ does not depend on $L_\parallel$ in this case,
from Eqs.~(\ref{eq:BW_chi}), (\ref{eq:leung_chi}) and
(\ref{eq:leung_chiwo}), we have three different predictions for the
susceptibility at $T_c$
\begin{eqnarray}
\mbox{Binder \& Wang}\qquad & \chi & \sim L_\perp^3,
\label{eq:chiLc} \\
\mbox{Leung (with $u$)}\qquad & \chi & \sim L_\perp^{7/3}, 
\label{eq:chiLb} \\
\mbox{Leung (without $u$)}\qquad & \chi & \sim L_\perp^2.
\label{eq:chiLa}
\end{eqnarray}
We will use Eqs.~(\ref{eq:BW_chi})-(\ref{eq:chiLa}) to check
which version of the predictions for scaling describes the data
of computer simulation better.

\section{Anisotropic scaling results} 
\label{sec:res}
\begin{figure}[htp]
\epsfig{figure=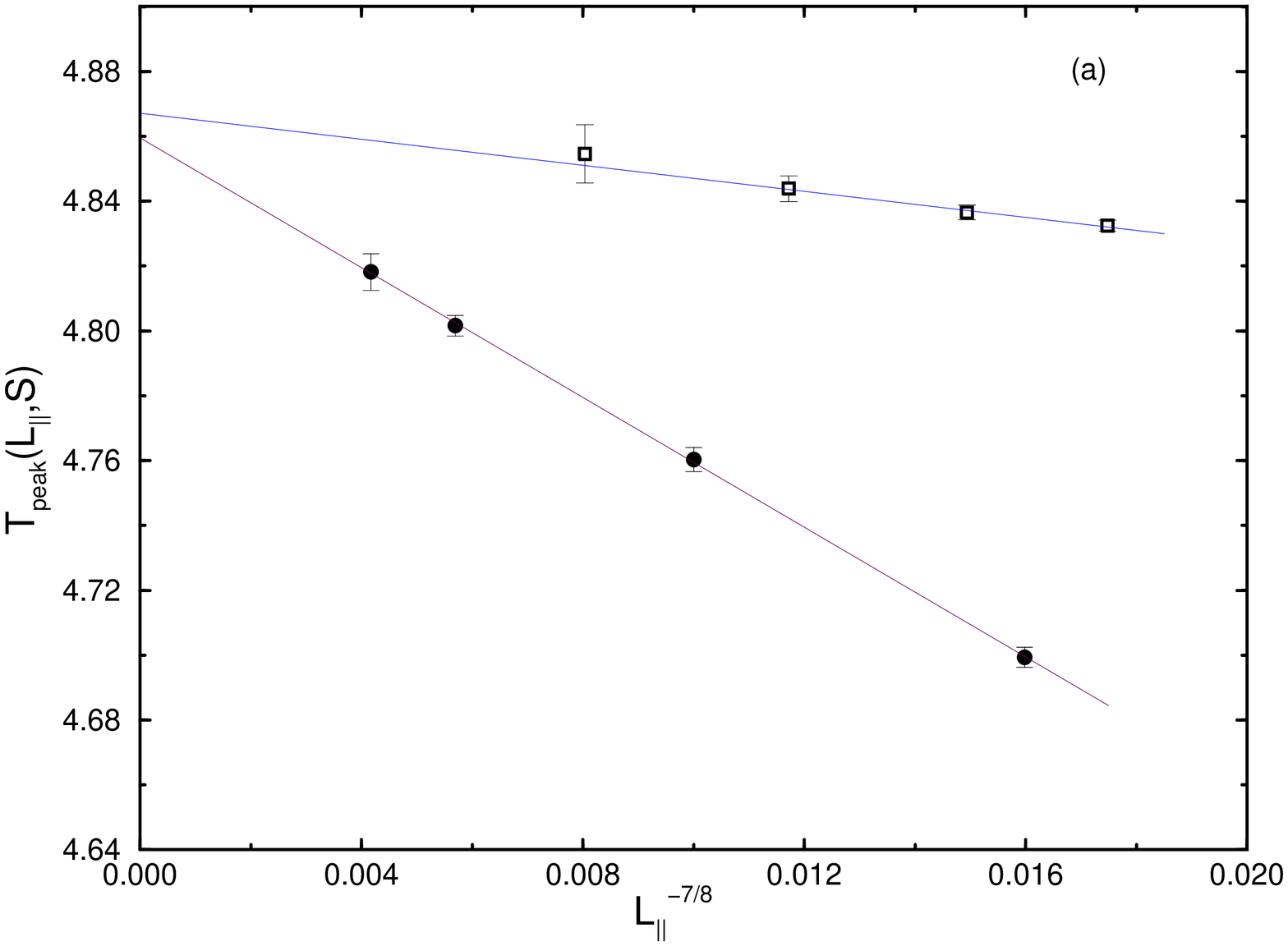,height=3.2in,angle=-90,
        bbllx=10bp, bblly=10bp, bburx=612bp, bbury=702bp}
\epsfig{figure=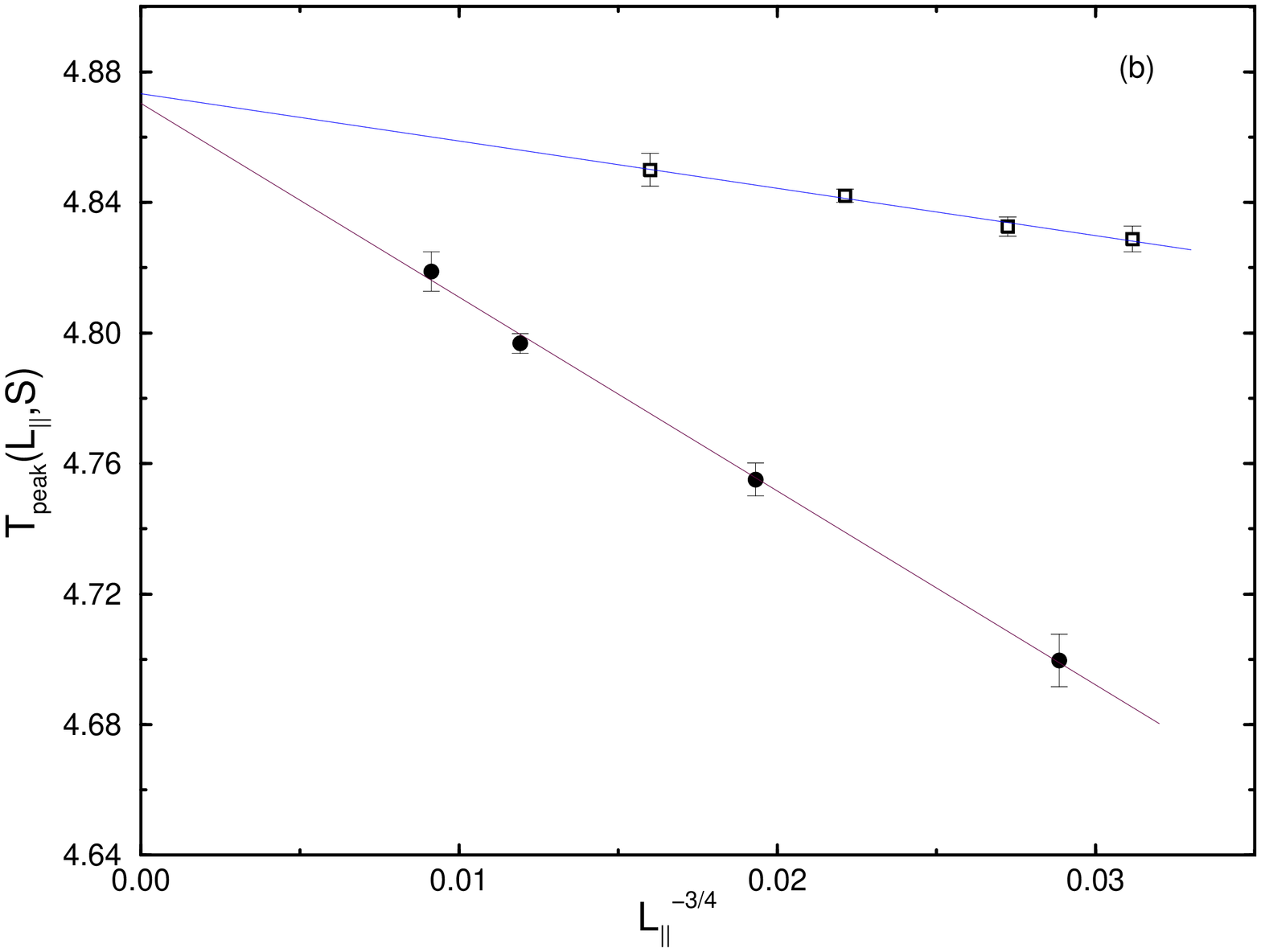,height=3.2in,angle=-90,
        bbllx=10bp, bblly=10bp, bburx=612bp, bbury=702bp}
\caption{The susceptibility peak location 
$T_{\rm peak}(L_\parallel, S)$  as a function of 
$L_\parallel^{-b}$; (a) $b=7/8$, (b) $b=3/4$.  
The system sizes $(L_\parallel,L_\perp)$ are 
(113,18), (193,22), (367,28), (524,32) [solid circle];
and (102,43), (122,46), (161,51), (248,60) [open square].
The limiting value
$T_{\rm peak}(L_\parallel\to \infty, S)$ is an estimate of $T_c$.}
\label{fig:Tc}
\end{figure}

An accurate determination of the critical temperature $T_c$ is
important for a quantitative analysis of the critical behavior.
We determine $T_c$ by the finite-size effect of the location $T_{\rm peak}$
of the susceptibility peak in $\chi$.   The data near peaks are 
fitted with a parabola to derive their heights and locations.
First we consider the
predictions in Eqs.~(\ref{eq:leung_chi}) and
(\ref{eq:leung_chiwo}).  Following well-known argument, the
shift of $T_c$ due to finite sizes is
\begin{equation}
T_{\rm peak}(L_\parallel, S) = T_c + a(S) L_\parallel^{-b},
\label{eq:Tc}
\end{equation}
where $a(S)$ is a scaling function, $b=7/8$ or $3/4$.  When
$S=L_\parallel^{3/8}/L_\perp$ is fixed, we have the usual shift
of the peak locations.  Fig.~\ref{fig:Tc}(a) shows 
$T_{\rm peak}$ against $L_\parallel^{-7/8}$, for two sets of data with
$S=0.327$ (first set, lower part) and $S=0.1317$ (second set,
upper part), respectively.  Least-square fit extrapolates to
critical temperatures $4.859\pm 0.005$ (first set) and $4.869
\pm 0.005$ (second set).  If the exponent $3/4$ is used in
Eq.~(\ref{eq:Tc}), according to Eq.~(\ref{eq:leung_chiwo}), the
estimate shifts to higher values of $4.870 \pm 0.005$ (first
set) $4.873\pm 0.005$ (second set).  Thus, while the
extrapolated $T_c$ from the two sets of data agree within errors
for both versions, the one with the exponent $3/4$ without the
dangerous irrelevant variable correction appears to be
marginally more consistent.  The consistency in $T_c$ for two
different sets of data with fixed $S$ is a significant
confirmation that $\lambda=8/3$ is the correct anisotropic
scaling exponent.

\begin{figure}[htp]
\epsfig{figure=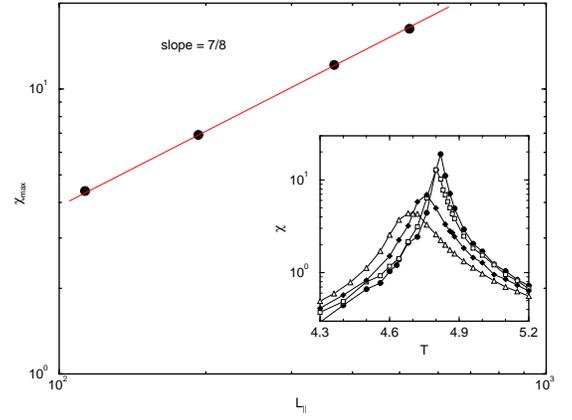,height=3.2in,angle=-90,
        bbllx=10bp, bblly=10bp, bburx=612bp, bbury=702bp}
\caption{Scaling of the susceptibility maxima.
The slope is as predicted in Eq.~(\ref{eq:leung_chi}),
within error from the best fit 0.86.
The insert shows the shift of the susceptibility peaks
for $(L_\parallel,L_\perp)$ being
(113,18), (193,22), (367,28), and (524,32) from left to right. 
}
\label{fig:xmax}
\end{figure}

The peak heights are additional information which we can
use.  According to Eq.~(\ref{eq:leung_chi}) and
(\ref{eq:leung_chiwo}), the peak height scales with system
size as $\chi_{\rm max} \sim L_\parallel^{7/8}$ or $\chi_{\rm max}\sim
L_\parallel^{3/4}$ if the variable $u$ is or is not taken into
account.  The nice feature of the susceptibility maximum
scaling is that it does not depend on the choice of the
second scaling variable $t \, L_\parallel^c$ in the scaling
functions.  In Fig.~\ref{fig:xmax}, we plot the height vs.
$L_\parallel$ in logarithmic scale.  The insert shows shift
of the susceptibility peaks.  Very nice linear behavior with
a slope of $0.86 \pm 0.02$ is obtained.  This is a solid
confirmation of Eq.~(\ref{eq:leung_chi}).

\begin{figure}[htp]
\epsfig{figure=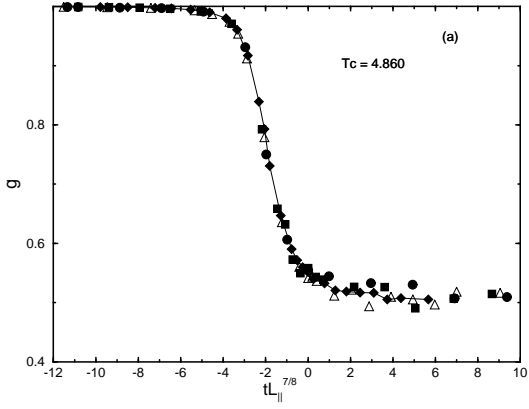,height=3.2in,angle=-90,
        bbllx=10bp, bblly=10bp, bburx=612bp, bbury=702bp}
\epsfig{figure=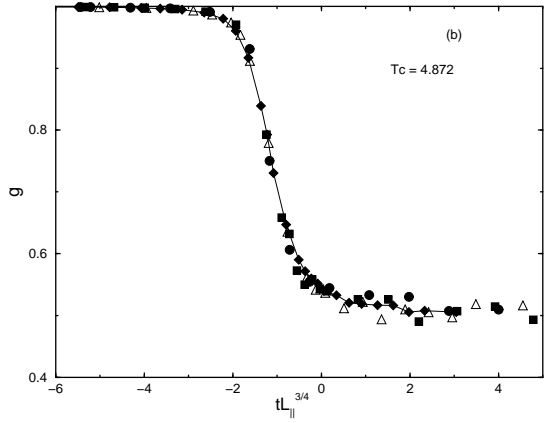,height=3.2in,angle=-90,
        bbllx=10bp, bblly=10bp, bburx=612bp, bbury=702bp}
\caption{The finite-size scaling of the fourth order cumulant $g$ with
the scaling variable 
(a) $t L_\parallel^{7/8}$ and $T_c = 4.860$; and
(b) $t L_\parallel^{3/4}$ and $T_c = 4.872$. 
The system sizes $(L_\parallel,L_\perp)$ are 
(113,18) $[$diamond$]$, (193,22) $[$triangle$]$, 
(367,28) $[$square$]$, (524,32) $[$circle$]$. }
\label{fig:g}
\end{figure}

For fixed $S$, the anisotropic finite-size scaling has the same
form as the isotropic scaling, involving a prefactor in $L$ and
just one scaling variable of the form $t\,L^c$.  We first look
at the scaling of the fourth order cumulant.  This quantity has
a simple scaling since there is no prefactor.
Fig.~\ref{fig:g}(a) is the cumulant $g$ plotted against scaling
variable $tL_\parallel^{7/8}$.  If the dangerous irrelevant
variable $u$ was ignored, the exponent $c$ would be $3/4$.  The
corresponding scaling plot is Fig.~\ref{fig:g}(b).  Both of them
yield equally well data collapsing if different $T_c$ are used.

\begin{figure}[htp]
\epsfig{figure=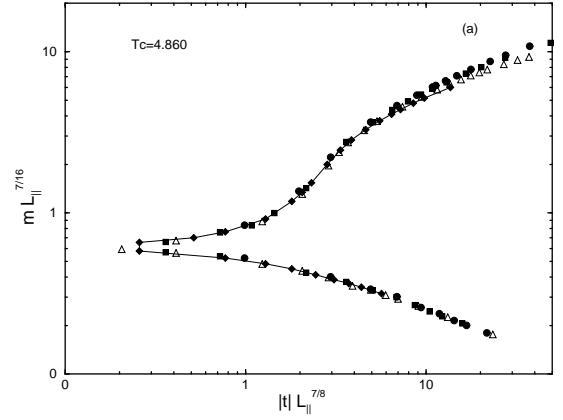,height=3.2in,angle=-90,
        bbllx=10bp, bblly=10bp, bburx=612bp, bbury=702bp}
\epsfig{figure=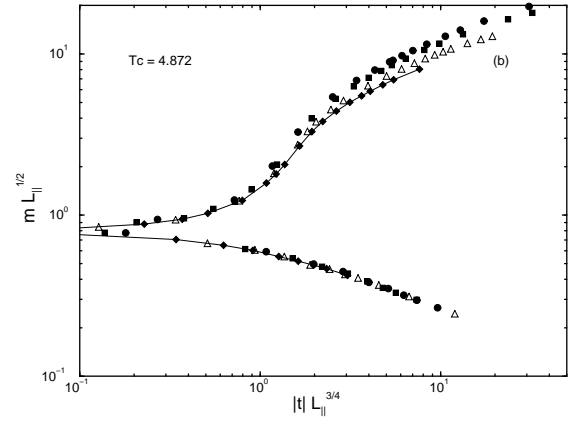,height=3.2in,angle=-90,
        bbllx=10bp, bblly=10bp, bburx=612bp, bbury=702bp}
\caption{The finite-size scaling of the order parameter $m$ with
the scaling variable 
(a) $t L_\parallel^{7/8}$ and $T_c = 4.860$; and
(b) $t L_\parallel^{3/4}$ and $T_c = 4.872$.
The system sizes are the same as in the previous figure.}
\label{fig:m}
\end{figure}

The finite-size scaling of the order parameter is presented in
Fig.~\ref{fig:m}.  Excellent scaling is found there when $u$ is
taken into account, using Eq.~(\ref{eq:leung_m}).  The upper
branch is for $t<0$ and the lower branch is for $t>0$.
Fig.~3(a) shows that the upper branch has a slope of about 0.5
for large value of $tL_\parallel^{7/8}$, which is consistent
with the exponent $\beta = 1/2$.  If $u$ is ignored, using
Eq.~(\ref{eq:leung_mwo}) instead, Fig.~3(b) shows that the data
collapse is not as good below $T_c$.

\begin{figure}[htp]
\epsfig{figure=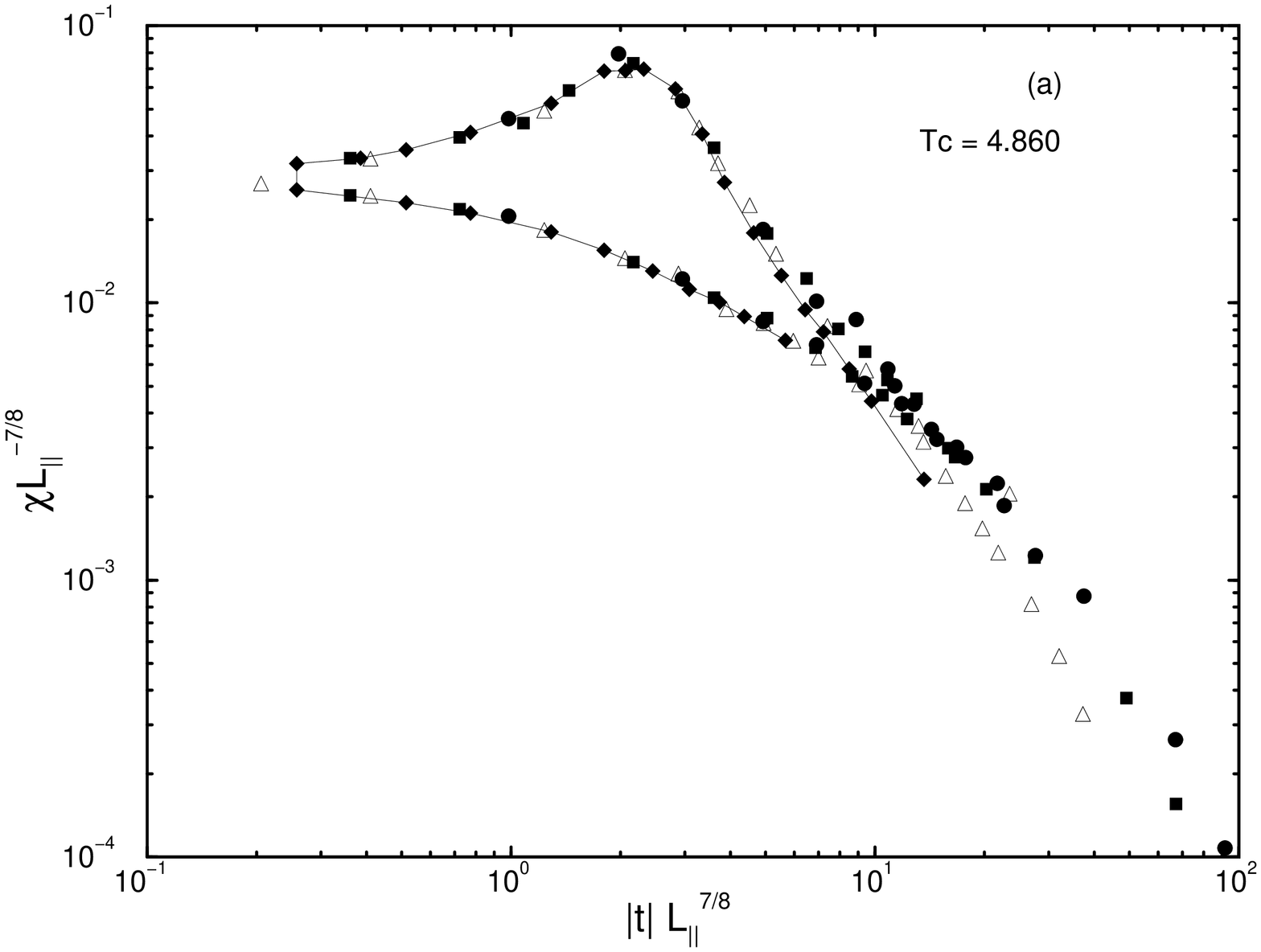,height=3.2in,angle=-90,
        bbllx=10bp, bblly=10bp, bburx=612bp, bbury=702bp}
\epsfig{figure=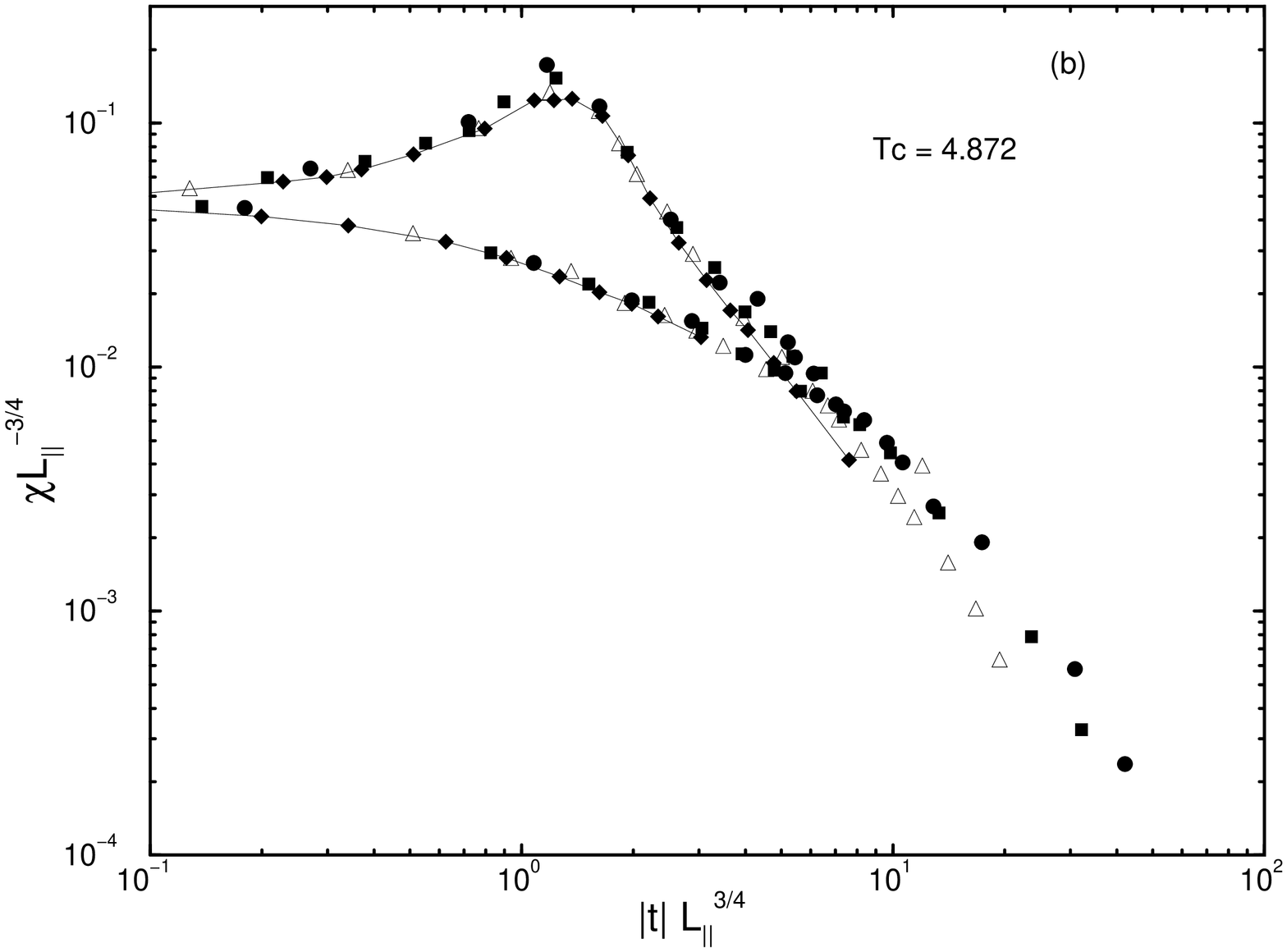,height=3.2in,angle=-90,
        bbllx=10bp, bblly=10bp, bburx=612bp, bbury=702bp}
\caption{The finite-size scaling of the susceptibility $\chi$ with
the scaling variable 
(a) $t L_\parallel^{7/8}$ and $T_c = 4.860$; and
(b) $t L_\parallel^{3/4}$ and $T_c = 4.872$.
The system sizes are the same as in the previous figure.}
\label{fig:x}
\end{figure}

The susceptibility data are shown in Fig.~\ref{fig:x} in scaling
form.  The upper curve with a peak is for $t<0$ and the lower
curve is for $t>0$.  The scaling is not as good as that for the
order parameter at low temperatures.  A plausible explanation
for such deviations from scaling is that the low-$T$ data may
fall outside the critical region.  If this is the case, the size
of the critical region could then be estimated to be about 10\%
of $T_c$. This interpretation is supported by the same kind of
deviation in the low-$T$ tails in $\chi$ and $m$ in equilibrium
Ising model\cite{cregion}, where the exponents are known exactly
and thus cannot be the source of deviations.  Similar behavior
is also found in two-dimensional DDS\cite{Wang96}.

\begin{figure}[htp]
\epsfig{figure=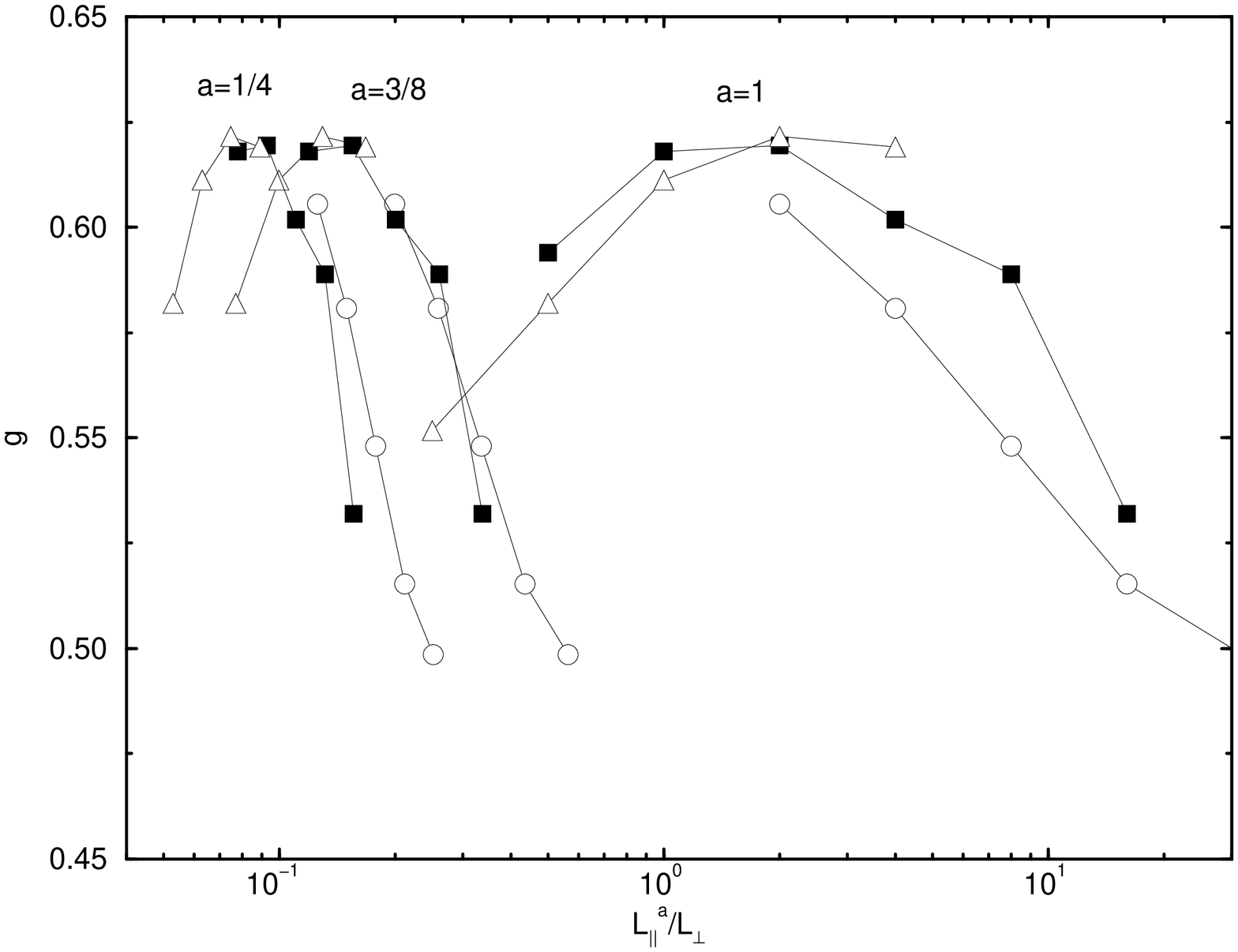,height=3.2in,angle=-90,
        bbllx=10bp, bblly=10bp, bburx=612bp, bbury=702bp}
\caption{The scaling of the fourth-order cumulant 
at $T_c = 4.860$.  The left set of curves with $a=1/4$ refers to
the prediction of Binder and Wang; the middle set with $a=3/8$
is that of Leung; the right set assumes isotropic scaling $a=1$.
Same symbol means the same transverse system size $L_\perp=$40,
30 and 20, from left to right.}
\label{fig:gT86}
\end{figure}

We now turn our attention to finite-size scaling at $T_c$, where
only one scaling variable of the form
$L_\parallel^{1/\lambda}/L_\perp$ is left.  We simulate a wide
range of system sizes and shapes, no longer restricting to ones
with fixed $L_\parallel^{3/8}/L_\perp$.  The exponent $\lambda$
can be determined when ensemble averages for different
$(L_\parallel,L_\perp)$ fall on one curve when plotted against
$L_\parallel^{1/\lambda}/L_\perp$.  Since large systems take
very long time to equilibrate at $T_c$, we do not have very
precise data.  Nevertheless, it is sufficient to distinguish
among alternative predicted scaling forms.  First, the
fourth-order cumulant at $T_c=4.860$ is presented in
Fig.~\ref{fig:gT86}.  The prediction of Leung with or without
$u$ term has the same scaling variable
$L_\parallel^{3/8}/L_\perp$, while that of Binder and Wang is
$L_\parallel^{1/4}/L_\perp$.  Isotropic scaling variable is
$L_\parallel/L_\perp$. The three sets of curves correspond to
these three cases.  Leung's scaling appears better.  This plot
also shows that for $S \approx 0.14$ we get a maximum value in
$g$.  This value corresponds to geometries where two correlation
lengths have the same ratio to the respective linear dimensions.
For the assumption of $L_\parallel/L_\perp$ as scaling variable,
the data clearly do not scale well.  This appears to be a strong
evidence in favor of anisotropic scaling.  The same type of
plots assuming $T_c = 4.872$ does not give good scaling for all
the choices of the scaling variables.  This seems to imply that
$T_c = 4.860$ is a better estimate for the critical temperature.

\begin{figure}[htp]
\epsfig{figure=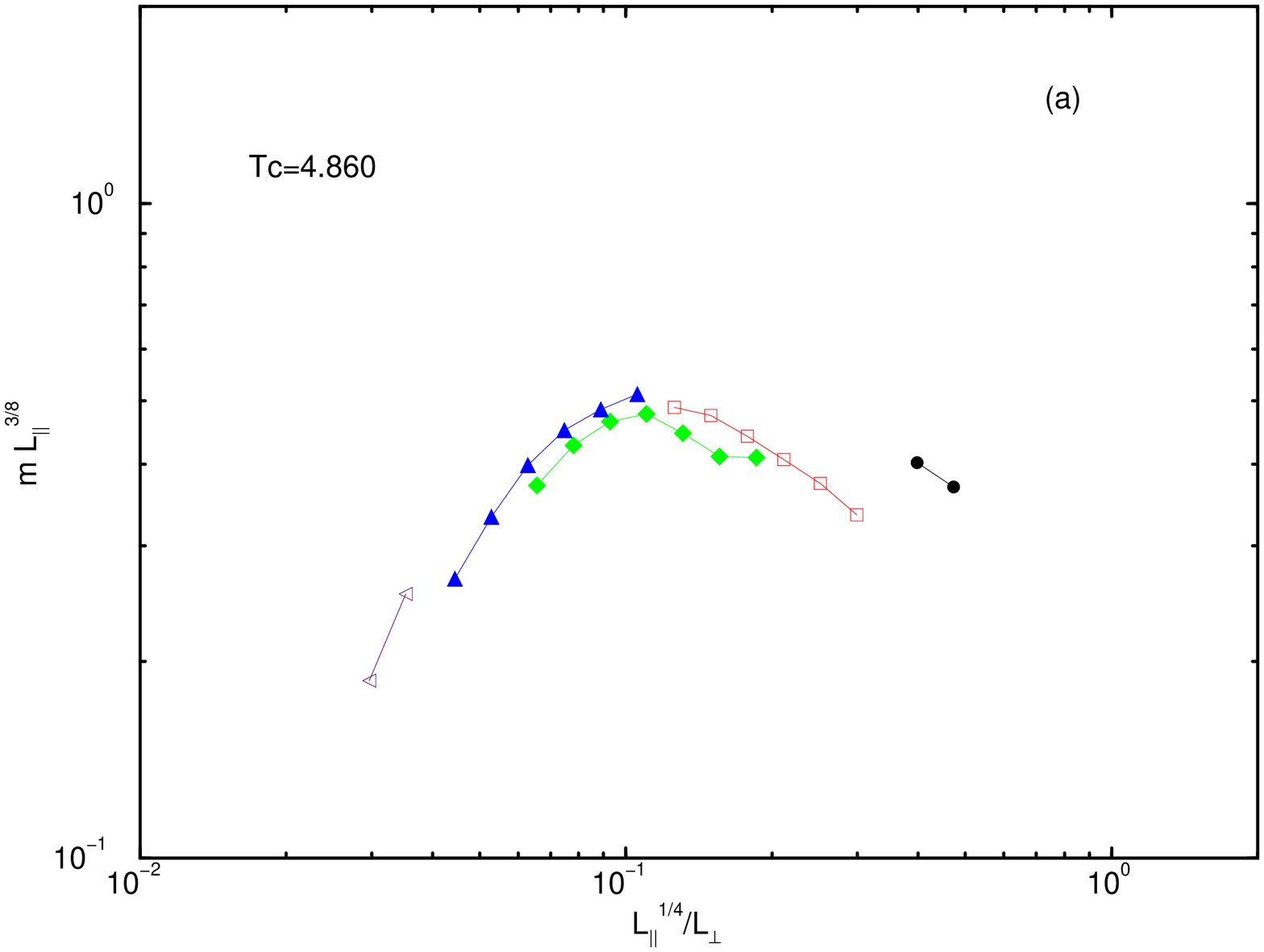,height=2.8in,angle=-90,
        bbllx=10bp, bblly=10bp, bburx=612bp, bbury=702bp}
\epsfig{figure=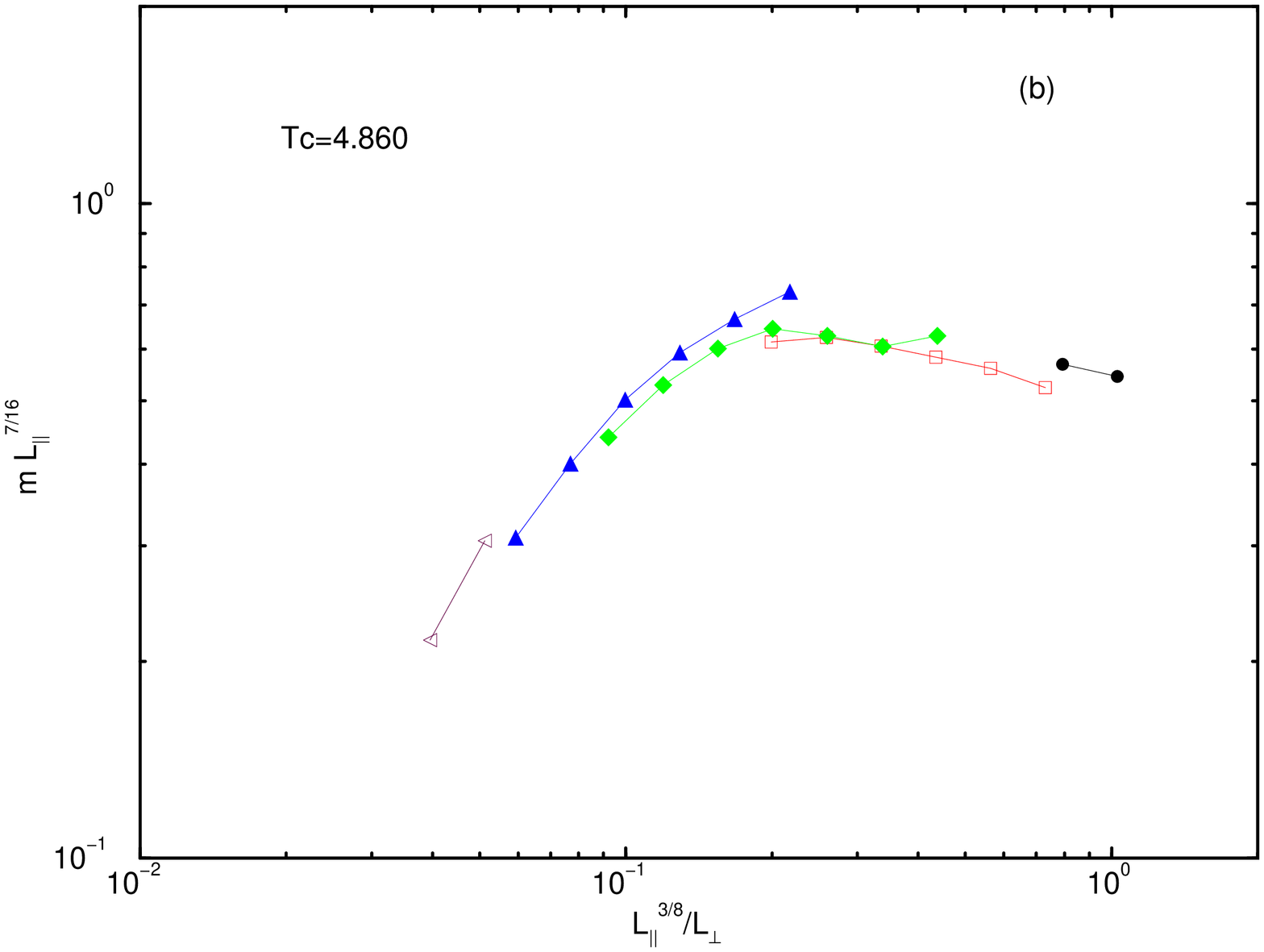,height=2.8in,angle=-90,
        bbllx=10bp, bblly=10bp, bburx=612bp, bbury=702bp}
\epsfig{figure=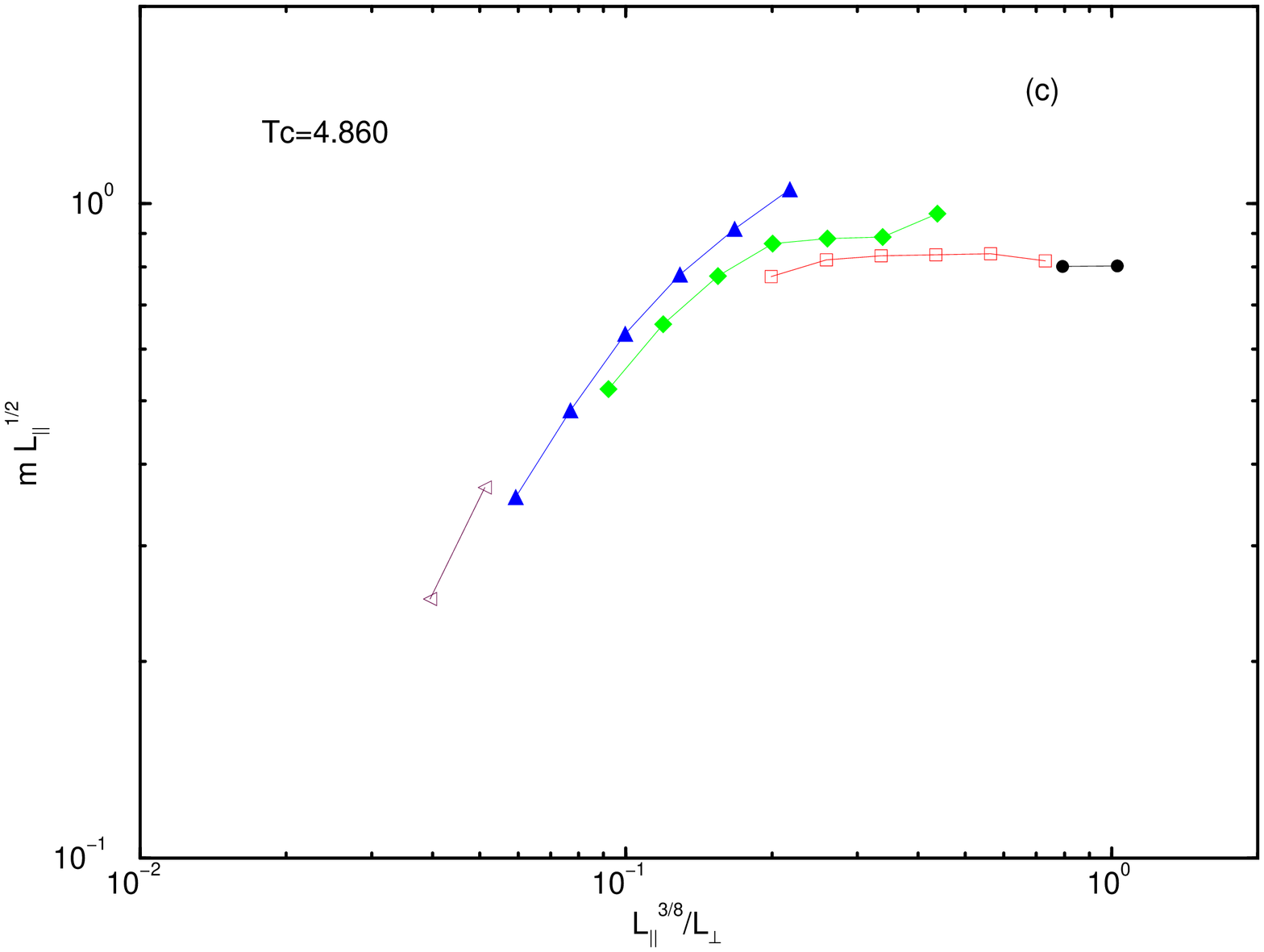,height=2.8in,angle=-90,
        bbllx=10bp, bblly=10bp, bburx=612bp, bbury=702bp}
\caption{The finite-size scaling of magnetization at $Tc=4.86$ for
(a) Binder and Wang scaling; 
(b) Leung scaling with $u$;
(c) Leung scaling without $u$.
Same symbol means the same transverse system size 
$L_\perp=$  60, 40, 30, 20 and 10, from left to right.}
\label{fig:mT86}
\end{figure}

In Fig.~\ref{fig:mT86} we show the scaling of magnetization at
$T_c=4.860$.  Fig.~\ref{fig:mT86}(a) uses Binder and Wang
scaling, (b) Leung's scaling with $u$ correction, and (c)
without $u$ correction.  Case (a) has large deviations;
case (b) generally scales better except at large
$L_\parallel^{3/8}/L_\perp$.  The scaling assuming $T_c=4.872$
(not shown) looks worse for case (a), and better for (c) at
large scaling variable.  Case (b) scales the best in both
temperatures.  This shows that the relative quality of scaling
is not sensitive to the choice of $T_c$ within its extrapolated
range.  The scaling function $\tilde m(x)$ has the following
asymptotic behaviors at large $L_\parallel$ or $L_\perp$ limit
due to the sum of magnetization of totally independent regions.
For small $x$, $m \propto 1/L_\perp$, this implies $\tilde m(x)
\sim x$ for all three cases.  for large $x$, we have $ m \propto
1/L^{1/2}_\parallel$, thus $m(x)
\sim x^y$, $y=-1/2$, $-1/6$, 0, respectively for case (a), (b), and (c).   
Unfortunately, all three plots are more or less consistent with this 
asymptotic slopes and thus alone it cannot give a sensitive test. 

\begin{figure}[htp]
\epsfig{figure=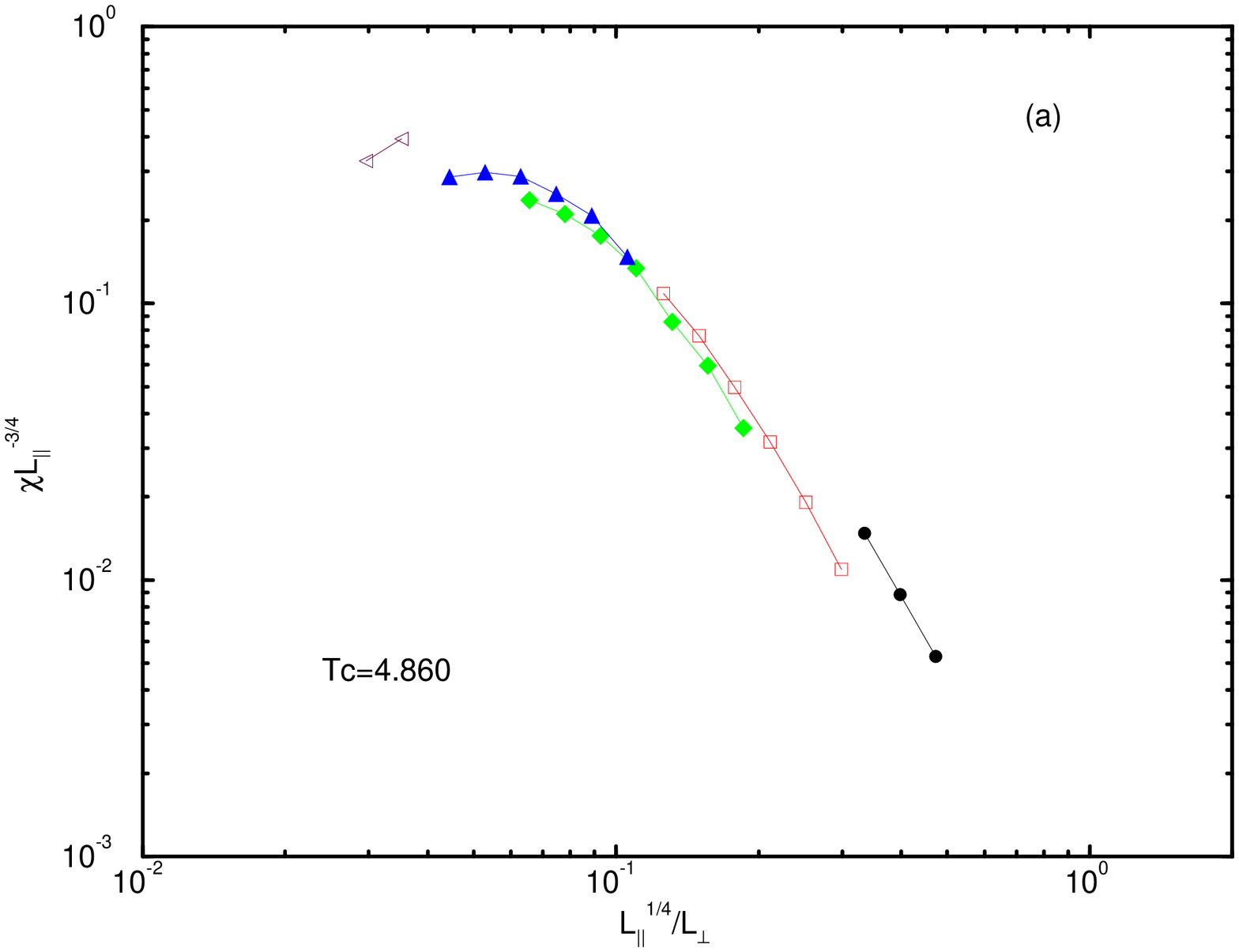,height=2.8in,angle=-90,
        bbllx=10bp, bblly=10bp, bburx=612bp, bbury=702bp}
\epsfig{figure=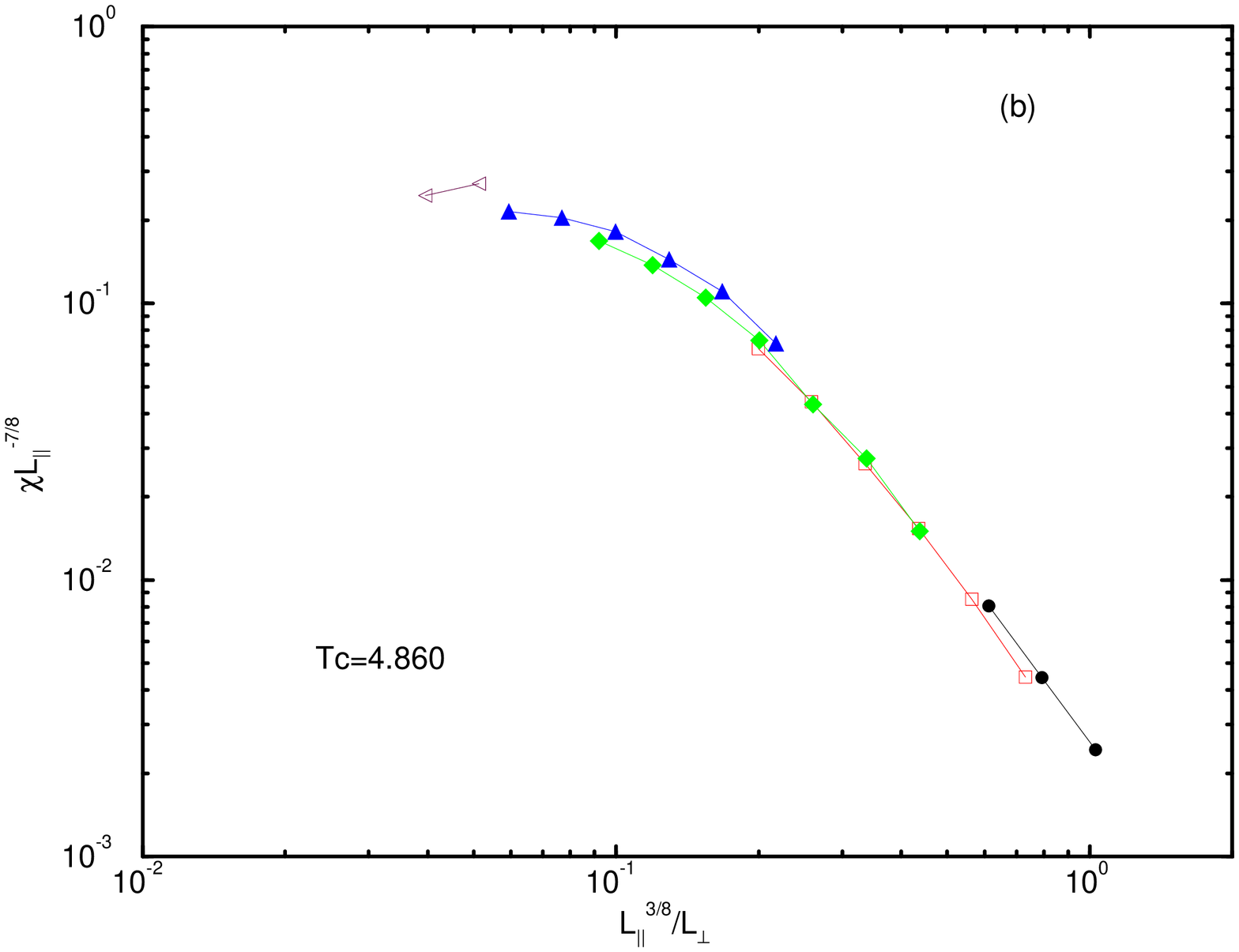,height=2.8in,angle=-90,
        bbllx=10bp, bblly=10bp, bburx=612bp, bbury=702bp}
\epsfig{figure=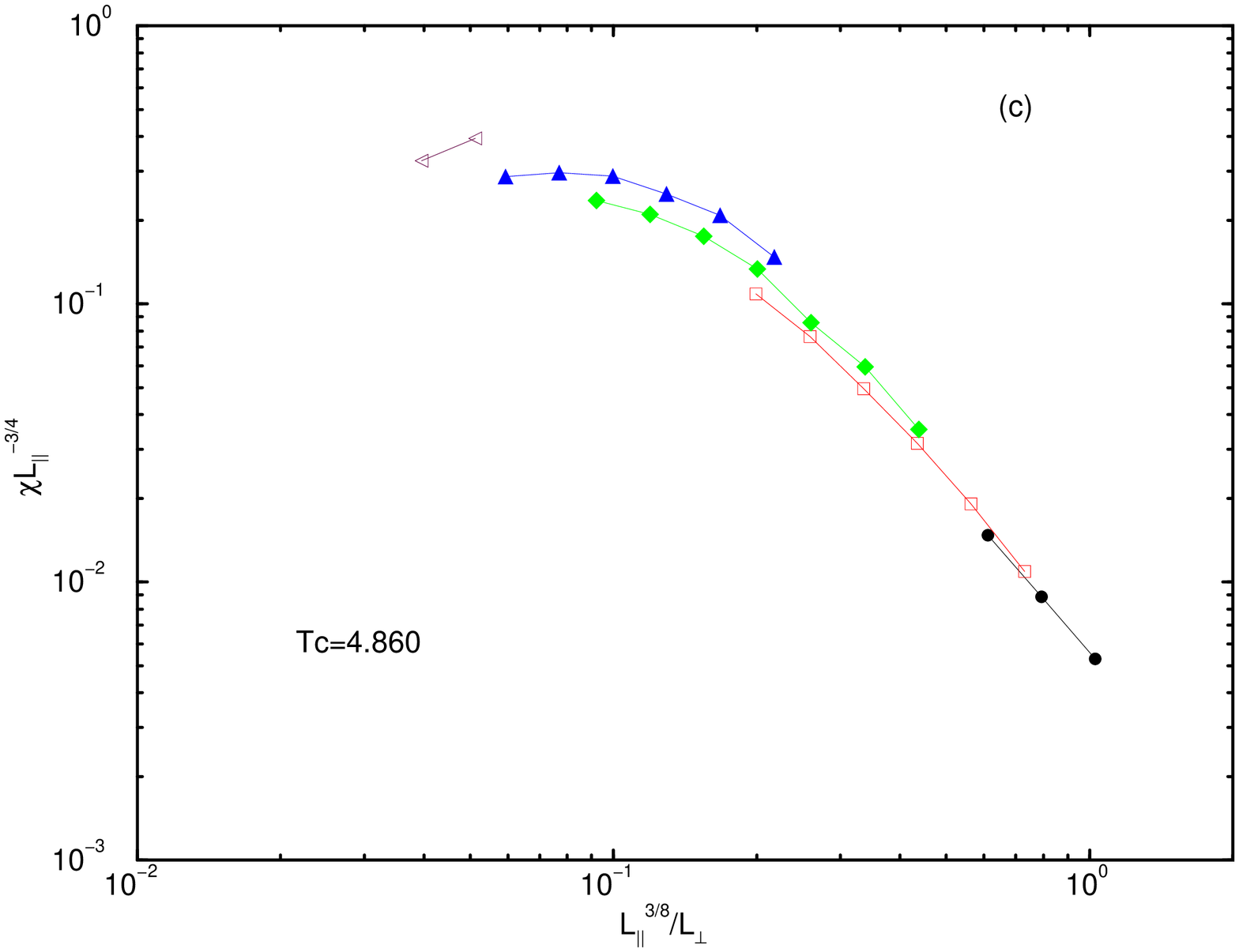,height=2.8in,angle=-90,
        bbllx=10bp, bblly=10bp, bburx=612bp, bbury=702bp}
\caption{The finite-size scaling of susceptibility at $Tc=4.86$ for
(a) Binder and Wang scaling; 
(b) Leung scaling with $u$;
(c) Leung scaling without $u$.
Same symbol means the same transverse system size 
$L_\perp=$  60, 40, 30, 20 and 10, from left to right.}
\label{fig:chiT86}
\end{figure}

Fig.~\ref{fig:chiT86} show the corresponding scaling plots for
the susceptibility at the same choice of $T_c$.  The trend is
the same as in $m$.  Although all three cases satisfy
Eqs.~(\ref{eq:chiLc})--(\ref{eq:chiLa}) for large
$L_\parallel^{1/\lambda}/L_\perp$, case (b) is most consistent
with the notion of a scaling function $\tilde \chi(x)$.

\begin{figure}[htp]
\epsfig{figure=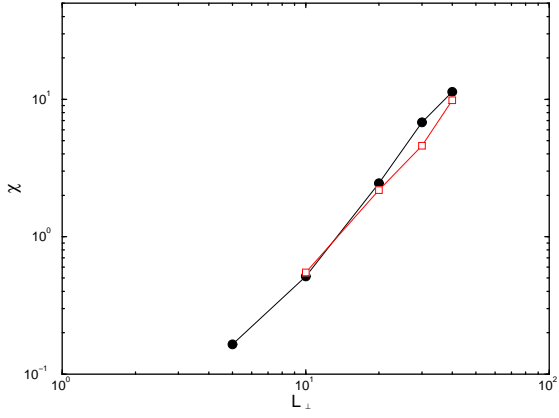,height=3.2in,angle=-90,
        bbllx=10bp, bblly=10bp, bburx=612bp, bbury=702bp}
\caption{The susceptibility $\chi$ at $T_c$ 
in the limit $L_\parallel \to \infty$ as a function of the
perpendicular dimension $L_\perp$ in logarithmic scales. The
critical temperature used is $T_c = 4.860$ for the circles and
$T_c=4.872$ for open squares.  }
\label{fig:chiL}
\end{figure}

Finally, in Fig.~\ref{fig:chiL}, we do a separate test of the
predictions of Eqs.~(\ref{eq:chiLc})--(\ref{eq:chiLa}) for the very
long geometry ($L_\parallel \to \infty$, $L_\perp$ finite).  The
critical temperature is taken to be $T_c \approx 4.860$.  The
$L_\parallel \to \infty$ limit is obtained by systems (1280,20),
(960,30), (320,40) together with smaller systems and
extrapolated to large $L_\parallel$, assuming a $1/L_\parallel$
convergence or similar power.  The least-square fit to the data
in Fig.~\ref{fig:chiL} gives exponent $2.27\pm 0.07$, in good
agreement with Leung's scaling with $u$ taken into account.
However, if $T_c=4.872$ is used, the exponent reduces to
$2.04\pm 0.09$.  Thus the distinction with or without
$u$-correction is not clear cut.  On the other hand, Binder and
Wang's scaling requires the exponent to be 3.  This seems
unlikely to be satisfied, due to the lack of data collapse in
Fig.~\ref{fig:chiT86}(a).  This demonstrates that to pass a
consistency check at $T_c$ requires both a good data collapse
and the correct asymptotic behaviors.

\section{Isotropic scaling} 

\begin{figure}[htp]
\epsfig{figure=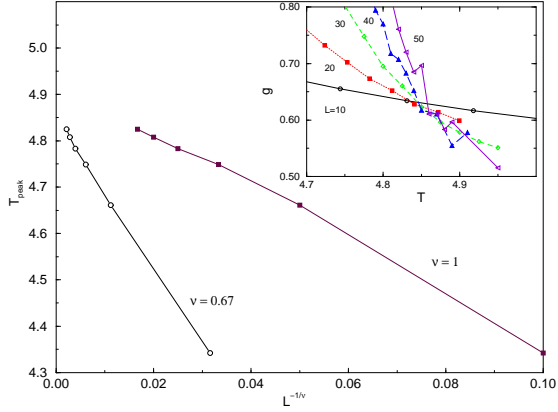,height=3.2in,angle=-90,
        bbllx=10bp, bblly=10bp, bburx=612bp, bbury=702bp}
\caption{The location of the susceptibility peak v.s. $L^{-1/\nu}$ 
for $\nu= 0.67$ and 1.  The insert shows the intersections of
the fourth order cumulant $g$ between different $L$.}
\label{fig:cb-Tc}
\end{figure}

There have been arguments \cite{Santos} that DDS 
under infinitely large driving field could follow the
normal isotropic finite-size scaling with one single correlation
length exponent, although no specific prediction of scaling exponents
are given.  For completeness, we test it by analyzing cubic systems 
with $L=20$, 30, 40, 50, and 60, assuming the usual finite-size scaling:
\begin{equation}
\chi(T) = L^{\gamma/\nu} \tilde \chi(t L^{1/\nu}),
\label{eq:iso_chi}
\end{equation}
\begin{equation}
m(T) = L^{-\beta/\nu} \tilde m(t L^{1/\nu}),
\label{eq:iso_m}
\end{equation}
\begin{equation}
g(T) = \tilde g(t L^{1/\nu}).
\label{eq:iso_g}
\end{equation}
A second scaling variable $L_\parallel/L_\perp$, equal to one in our data,
is not explicitly written.
In Fig.~\ref{fig:cb-Tc}, we plot the location $T_{\rm peak}$ of the
susceptibility peak versus $L^{-1/\nu}$ with two choices of $\nu$.  
Values of $\nu$ outside this range result in unacceptable,
nonlinear behavior.  
This plot therefore gives us an idea of the bounds of $T_c$.
Peak extrapolation depends on the assumption on
$\nu$, varying from 4.86 to 4.90 for $\nu = 0.67$ to 1. 
The curve is not quite linear for any choice of $\nu$; 
this is expected if the true behavior is given by Eq.~(\ref{eq:Tc}).  
The insert shows the intersections of the fourth-order
cumulant.  The intersections shift towards higher values from
4.83 to 4.86 as $L$ increases.  If we trust the larger system sizes, 
the estimate of $T_c$ would be 4.86.  This is the value barely 
in agreement with the extrapolation.

\begin{figure}[htp]
\epsfig{figure=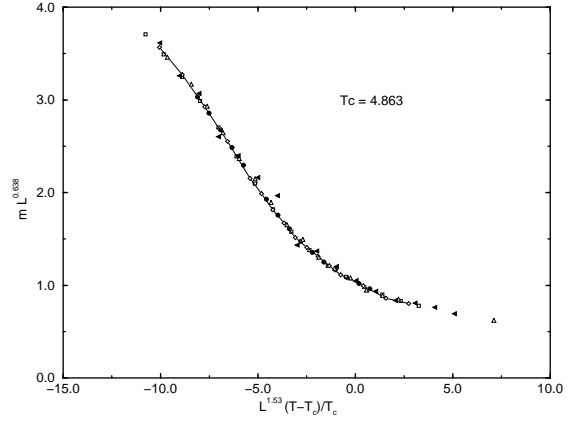,height=3.2in,angle=-90,
        bbllx=10bp, bblly=10bp, bburx=612bp, bbury=702bp}
\caption{Scaling of the magnetization for the cubic systems.  
The linear sizes are 20 (solid circle), 30 (square), 40
(diamond), 50 (open triangle), 60 (solid triangle).}
\label{fig:cb-m}
\end{figure}

\begin{figure}[htp]
\epsfig{figure=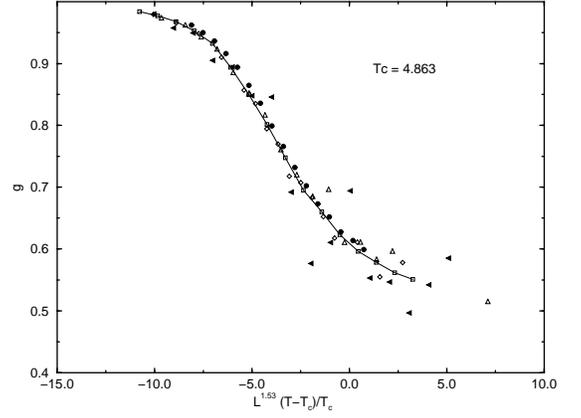,height=3.2in,angle=-90,
        bbllx=10bp, bblly=10bp, bburx=612bp, bbury=702bp}
\caption{Scaling of the fourth order cumulants for the cubic systems.  
The system sizes are the same as in the previous figure.}
\label{fig:cb-g}
\end{figure}

\begin{figure}[htp]
\epsfig{figure=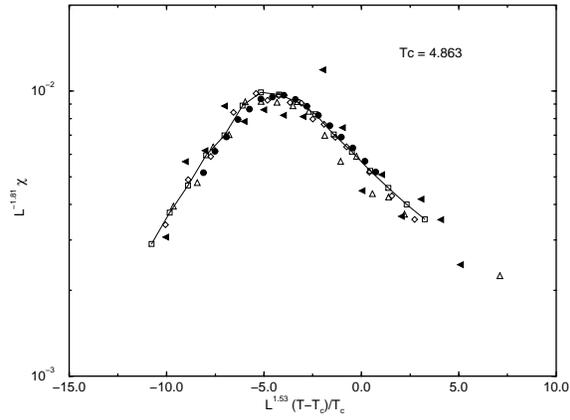,height=3.2in,angle=-90,
        bbllx=10bp, bblly=10bp, bburx=612bp, bbury=702bp}
\caption{Scaling of the susceptibility for the cubic systems.  
The system sizes are the same as in the previous figure.}
\label{fig:cb-x}
\end{figure}

Having determined $T_c$, 
the best off-$T_c$ scaling is seen in $m$,
achieved by choosing $T_c = 4.863$, $1/\nu=1.53$, 
and $\beta/\nu = 0.638$ (see Fig.~\ref{fig:cb-m}).  However, this
set of parameters is not the best choice for $g$.
Figure~\ref{fig:cb-g} shows that with the same $T_c$ and
$\nu$ as for $m$.  Due to the fact that $g$ curves are not
intersecting at a unique $T$, systematic deviations are observed
no matter what values of $T_c$ and $\nu$ are used.  
Next, we determine
$\gamma/\nu=1.81$ from the susceptibility peak height vs. system size.  
Relatively large deviation also occurs for the 
off-$T_c$ susceptibility data (Fig.~\ref{fig:cb-x}).  

\begin{figure}[htp]
\epsfig{figure=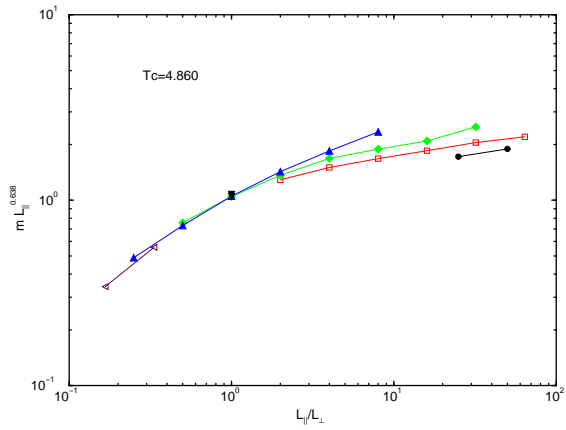,height=3.2in,angle=-90,
        bbllx=10bp, bblly=10bp, bburx=612bp, bbury=702bp}
\caption{Isotropic scaling for magnetization at $T_c$, using 
the same set of original data as in Fig.~\ref{fig:mT86}.}
\label{fig:isom}
\end{figure}

Even though the data more or less scale in $t L^{1/\nu}$ with
a suitable set of exponents, we see systematic deviations and
slight inconsistency among different quantities. 
The quality of data collapsing is inferior 
to that of the anisotropic scaling except perhaps for $m$.  
Since the critical temperature estimated from isotropic scaling 
agrees with that of anisotropic scaling,  the
same data obtained for the purpose of testing anisotropic scaling 
at $T_c$ can also be used to test isotropic scaling.
The results do not support the latter.
We have already seen that the fourth order cumulant $g$
does not scale (Fig.~\ref{fig:gT86}, $a=1$).  
In Fig.~\ref{fig:isom} we show $m L_\perp^{\beta/\nu}$ vs.
$L_\parallel/L_\perp$ at $T_c$. If the system is isotropic, 
it should scale.  But clearly, we see large systematic deviations.
Finally, in the quasi-one-dimensional limit, 
the equation analogous to Eqs.~(\ref{eq:chiLc})--(\ref{eq:chiLa})
is $ \chi \propto L_\perp^{\gamma/\nu} = L_\perp^{1.8}$, but
the actual exponent from the data is much larger (Fig.~\ref{fig:chiL}). 

The reason for the above good fits off $T_c$ and the poor fits at $T_c$
is a consequence of our fitting the former first.  This warrants
the crossing at $L_\parallel/L_\perp=1$ in Fig.~\ref{fig:isom}.
The fact that there is no collapse elsewhere 
along the $L_\parallel/L_\perp$ axis is a clear 
indication of the failure of the assumption of isotropic scaling.
Thus, the moral of this exercise is that 
an apparently good fit from a partial test may be dangerously misleading
in the case of scaling with two variables. 

\section{Conclusion}
\label{sec:conclusion}

We have performed a large-scale simulation for the standard
driven diffusive model.  We test the theoretical predictions
of finite-size scaling against numerical data, 
as carefully, critically and completely as we can. 
The task is to decide which of the various competing theories
is the most consistent with the data.
There have been four theoretical proposals, characterized by
different sets of finite-size scaling exponents.
Two kinds of scaling tests are performed: at and off the critical temperature.
Both are necessary because there are two scaling variables involved.
The two versions due to Leung, with and without the dangerous irrelevant
variable correction, fit the off-$T_c$ data almost equally well if
$T_c$ is adjusted accordingly. 
As to the data taken at $T_c$,
while they do not have enough quality for us to
perform a stringent test,
the scaling behavior and susceptibility peak heights clearly
favor the one including the dangerous irrelevant variable.  
Such tests also indicate that Binder and Wang scaling may not be valid.
Regarding the recent proposal based on isotropic scaling,
we find that the data can be cast in scaling forms with some 
effective exponents, but its theoretical basis is more conjectural, 
and the quality of scaling is poorer especially at $T_c$.
Thus, taking into account all aspects of scaling, we conclude that
the field-theoretic prediction based on 
spatial anisotropies with dangerous irrelevant variable corrections is
the most satisfactory.

\section*{Acknowledgment}

The work of JSW was supported in part by an Academic Research
Grant No.~RP950601.  KtL acknowledges supports by the National
Science Council of ROC through grants NSC87-2112-M-001-006 and
NSC88-2112-M-001-013, and a Main-theme grant from the Academia
Sinica.

\bibliographystyle{plain}

\clearpage



\end{document}